\documentstyle[12 pt,aasms4]{article}
\begin{document}
\title{Possible New Identifications for EGRET Sources}
\author{Steven D. Bloom \altaffilmark{1}and Robert C. Hartman}
\affil{Laboratory for High Energy Astrophysics, NASA/ Goddard Space
Flight Center, Greenbelt, MD 20771}
\altaffiltext{1}{NAS/NRC Resident Research Associate}
\author{Harri Ter\"asranta, Merja Tornikoski and Esko Valtaoja}
\affil{Mets\"ahovi Radio Research Station, Helsinki University of
Technology, Otakaari 5A, SF-02150 Espoo, Finland}
\begin{abstract}
The majority of the EGRET objects remain either unidentified or
questionably identified at other wavelengths. We have conducted
a multiwavelength study of several radio sources within or near the error
boxes of EGRET unidentified
sources at mid to high galactic latitude, under the hypothesis that the radio
sources are blazars and are thus the best identification candidates for
the EGRET objects. We show that one of these radio sources, PMN 0850-121,
has a flux of 1.5 Jy at 22 GHz, and a  nearly flat spectrum up to
230 GHz and is thus very likely to be the correct identification
for the EGRET source 2EG J0852-1237. 
\end{abstract}
\section{Introduction}
Though 59  EGRET sources have been identified as either blazars or
pulsars, there remain 95 sources which have not yet been identified
with objects at other wavelengths (\cite{tho96}).
\cite{mcl96} hypothesize that many of the nonvariable unidentified sources
are part of a broad galactic population, possibly pulsars. Due to EGRET's 
diminishing 
efficiency, we may not learn much more about the high energy gamma-rays 
(eg. positions or variability)
from these sources until the next mission.
Any further insights on these sources in the intervening time will most 
likely be gained at
other wavelengths.
We have therefore initiated a multiwavelength radio/millimeter-wave study in 
order to assess the credibility of possible counterparts.  
\cite{mat97} determine, from the
Green Bank Survey radio catalogues, which flat spectrum radio sources are 
within or near the error
boxes of each EGRET source. Based on the flux
density of the radio source at 4.85 GHz, the spectral index between
1.4 and 4.85 GHz, 
and the 
candidate source's position within the EGRET error ellipse, they
calculate an {\it a posteriori} probability that the candidate source
is a correct identification for the EGRET source. From their lists, we have extracted 4 sources, with relatively 
high probabilities for identification, and we have included them in our
multiwavelength study. We are working under the hypothesis that these
sources are blazars, and are the sources of the gamma-rays.
To our knowledge, these sources all previously only had non-simultaneous measurements
at the two frequencies of the Green Bank Survey (1.4 and 4.85 GHz), and no 
other observations except for non-simultaneous optical observations (by other
observers) to 
determine redshifts and magnitudes. 
Our main goal is
to obtain quasi-simultaneous centimeter to millimeter-wave spectra to 
determine whether
each flat spectrum radio source has blazar properties. Certain blazar
properties which are correlated with gamma-ray detection by EGRET, 
such as a flat spectrum down to millimeter wavelengths,
and/or significant variability, would lead to an even higher probability
for identification than that from the formal calculation ( \cite{mat97}).     
In section 2
we describe the observations, in section 3 we analyze the spectra
of each object and assess the possibilities of new identifications
and in section 4 we present our conclusions and discuss our future work.
\section{Observations}
The sources, positions. and flux densities at each frequency are listed in  Table 1.  Column (1) gives the name of the EGRET source; Column(2) the name of the
radio source that is a potential identification; Column(3) gives the redshift
if known; Column (4) gives the type of object, if known; Columns (5) and (6)
give the J2000 celestial coordinates; right ascension (RA) in hh mm ss;
declination (DEC) in dd mm ss. For most sources, this position is the radio
position as found by the NASA Extragalactic Database (NED), but
for PMN 0850-121 we have used the more accurate optical position (\cite{mat97};\cite{hal97}). Columns (7-12) give the averaged flux densities
(with typical uncertainties)
at the respective frequencies (in Jy); Column (13) gives the spectral index
between 2.25 and 22 GHz. For the sources that do not have data at 22 GHz, 
we have calculated the spectral index between 2.25 and 8.3 GHz.
In this table, and throughout this paper we use the convention 
$F_{\nu} \propto \nu^{\alpha}$.
 
We have used the 13.7 meter
antenna of Mets\"ahovi Radio Research Station (details of observing
methods can be found in
\cite{ter92})
to observe at 22 and 37 GHz as part of a regular monitoring program
of blazars and blazar candidates. 
The observations were conducted 20 Mar 1997 and 1 Apr 1997.
As part of this same program,
two of these sources (PMN 0850-121 and PKS 2320-035) were observed at 90 and 
PMN 0850-121 was observed at 230 GHz with the Swedish-ESO Submillimeter Telescope
(\cite{tor96}).
The remaining sources were either at declinations too high to be seen
from Chile, or were too close to the Sun.  These observations
were conducted on 1 Apr 1997 and again on 31 Apr/1 May.
Each of our sources was monitored regularly (usually more than once
per week) with the 2 element Green Bank Interferometer (one antenna of
the 3 element array is not used for monitoring studies)
at 2.25 and 8.3 GHz. Each antenna is 26 meters and the baseline is 
2400 meters. The observations were conducted between 25 Mar 1997 and
12 May 1997.
(see \cite{wal96} for details of the GBI). The centimeter--millimeter range was picked because the known
EGRET blazars have spectra which are nearly flat in this range (\cite{blo94})
and it is generally believed that a $\sim 1$ Jy source with a flat
centimeter to millimeter spectrum in the error box is a good candidate for identification.
The observations were not precisely simultaneous, but were all within
a few days of each other, with the 2.25 and 8.3 GHz observations 
continuing after the higher frequency observations were conducted. 
The gamma-ray observations (from a much earlier time frame) are summarized
in \cite{tho95} and \cite{tho96}.
\section{Discussion}
We find that these sources, when observed simultaneously between
2.25 and 22 GHz show a nearly flat or rising spectrum. This is an important
result, since previous estimates of spectral indices in this range
were based on measurements taken years apart and flat spectrum sources
are known to vary significantly at centimeter wavelengths over months to years.
The source PMN 0850-121 is brighter than 1 Jy from 22 to 230 GHz,
and is within the 95\% confidence contour of the test statistic map for the
associated gamma-ray source. By the standard of \cite{tho95},
this source would be considered a positive identifcation, although,
since the source is roughly 0.5 Jy at 5 GHz, the identification  probability, $p(id|r)$, of
\cite{mat97} is about 0.05, much lower than that of the 42 strong EGRET 
identifcations which he lists. On the other hand, $\alpha_{90-230}$ is -0.2, which is roughly similar to the
millimeter spectral inidces for EGRET sources as determined in
\cite{blo94}( $\langle \alpha \rangle $=-0.54), and considerablly flatter than 
$\alpha_{90-230}$ for similar blazars not detected by EGRET ($\langle \alpha
\rangle=-0.75$). Over the $\sim 1.5 $ months of GBI observations for
PMN 0850-121,
the 2.25 GHz data show a 10 \% increase and the 8.3 GHz show a 20 \% 
increase in flux density. 
Though the systematic uncertainties are estimated to be at roughly
the same level as this variability,
an analysis of several calibration
sources shows (\cite{wal97}) that this long term increase is not a systematic
effect. The smaller time-scale (and amplitude) increases seen throughout
the 2.25 GHz observations are likely caused by scinitillation (\cite{wal97}).
In addition, \cite{hal97} report that this source is highly
variable at optical wavelengths, with a flare of about 2 magnitudes 
from Feb 1996 to Feb 1997 (the time-scale of variation within that period is 
unkown). These data suggest that there may have been a small amplitude centimeter-wave 
flare following an earlier optical outburst by several months. 
Such activity has been seen in EGRET detected blazars, such as 
0836+710 (\cite{rei93}); however, we do not have enough data
to pin-point the relative amplitudes or times of the flares.

Considering its contemporaneously measured spectrum, the
formal value of $p(id|r)$ for PMN 0850-121 is likely to be an underestimate of
the true probability that this is the correct EGRET identification.
For the remaining sources, though their spectra above 22 GHz are not generally
known, it is clear that B1055+5644 and B2346+3832 have flat spectra up to this
point, and that PKS 2320-035 has a slowly decreasing spectrum up to 90 GHz. Though they are not extraordinarily bright, spectral
flatness up to 22 GHz  make them stronger candidates
for EGRET identifcation than the lower frequency data alone. 
Since many millimeter flares only propagate down to several GHz with a much
lower amplitude, and with delays of months, data at lower frequencies 
(less than 20 GHz) should only be used as a rough guidline for source
variability and possible identifcation with gamma-ray sources.  
PKS 2320-035
has a slightly higher flux density at 5 GHz (interpolated from measured
values) and a flatter spectral index, making $p(id|r)$  higher
than the value of \cite{mat97}, but still not quite as high as the identifications
considered to be firm. However, the observations at 90 GHz
show that the true probability for a positive identification
is likely to be higher. This source had continuously {\it
decreasing} flux density at 2.25 and 8.3 GHz by about 10 \% .
\section{Conclusions and Future Work}
We conclude that the source PMN 0850-121 is firmly identified with an EGRET source, and that
several other candidate sources from this study, though probably blazars, are too dim
to be considered {\it firm} identifications. However, future 
monitoring may reveal that these sources have higher centimeter/millimeter
states.
In addition to the total flux density data at centimeter to submillimeter
 wavelengths,
VLBI observations will also add to our knowledge of these sources.
In general, multiwavelength monitoring of candidate EGRET sources 
should be conducted on a wider scale to see whether more unidentified
sources at mid to high galactic latitudes are blazars.

We would like to thank J. Mattox for supplying us with the code
to determine identifcation probabilities using the Green Bank survey
data. We would also like to thank E. Waltman and F. Ghigo for
consultation on the GBI Data. The GBI is a facility of the National
Science Foundation operated by NRAO with support from the NASA
High Energy Astrophysics Program. This research has made use of
the NASA/IPAC Extragalactic Database (NED) which is operated by the
Jet Propulsion Laboatory, California Institute of Technology, under
contract with the National Aeronautics and Space Administration. 

\newpage
\centerline{\bf Figure Captions}
\figcaption[0850_var1.ps]{The 2.25 and 8.3 GHz data for PMN 0850-12. The
2 GHz data is represented by filled circles, and the 8 GHz data is represented
by open diamonds. The modified Julian day (JD-2400000.5) is for the period
March 25-12 May 1997}
\figcaption[1055_var1.ps]{The 2.25 and 8.3 GHz data for 1055+564}
\figcaption[2320_var1.ps]{The 2.25 and 8.3 GHz data for PKS 2320-035}
\figcaption[2346_var1.ps]{The 2.25 and 8.3 GHz data for 2346+383}
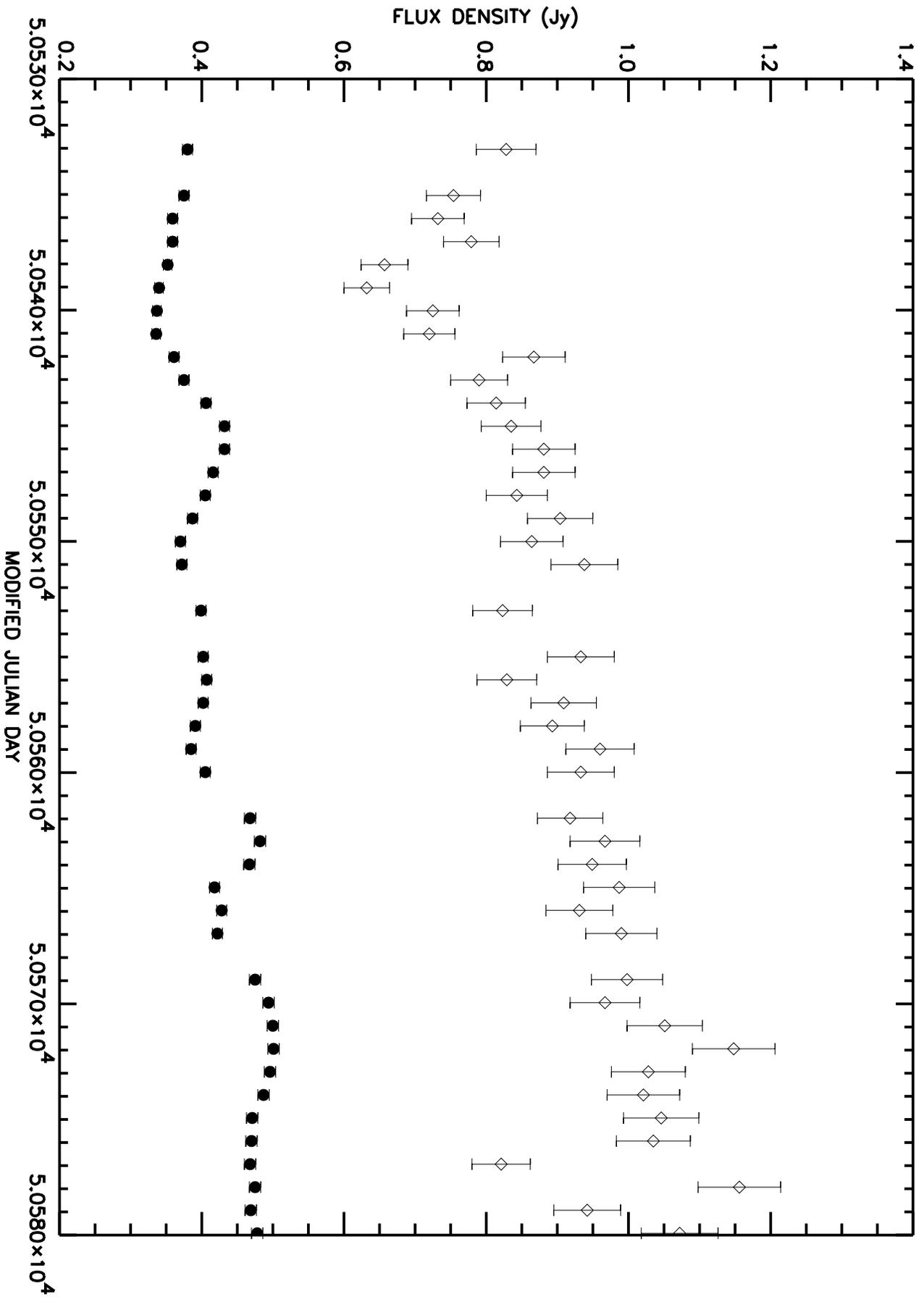
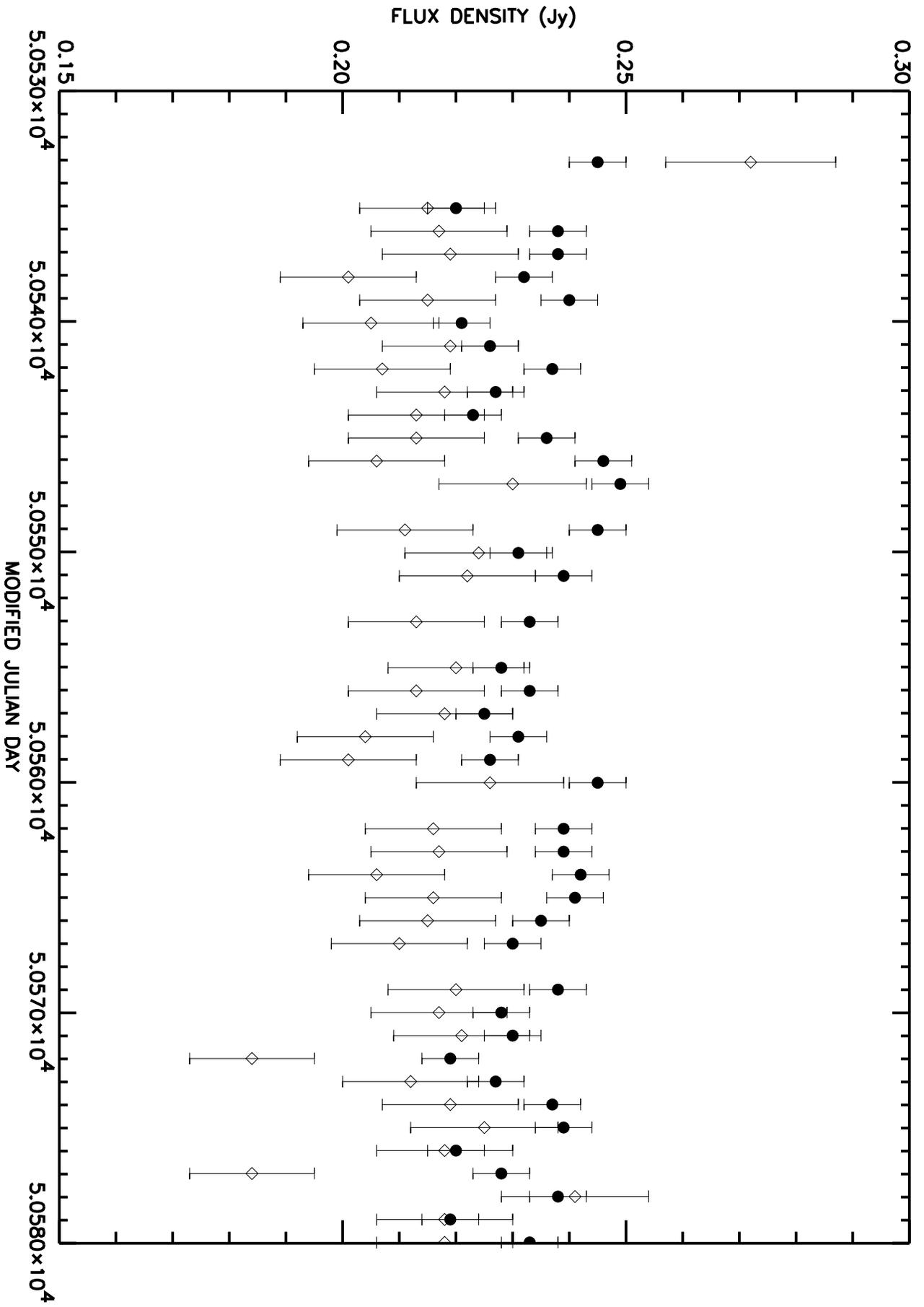
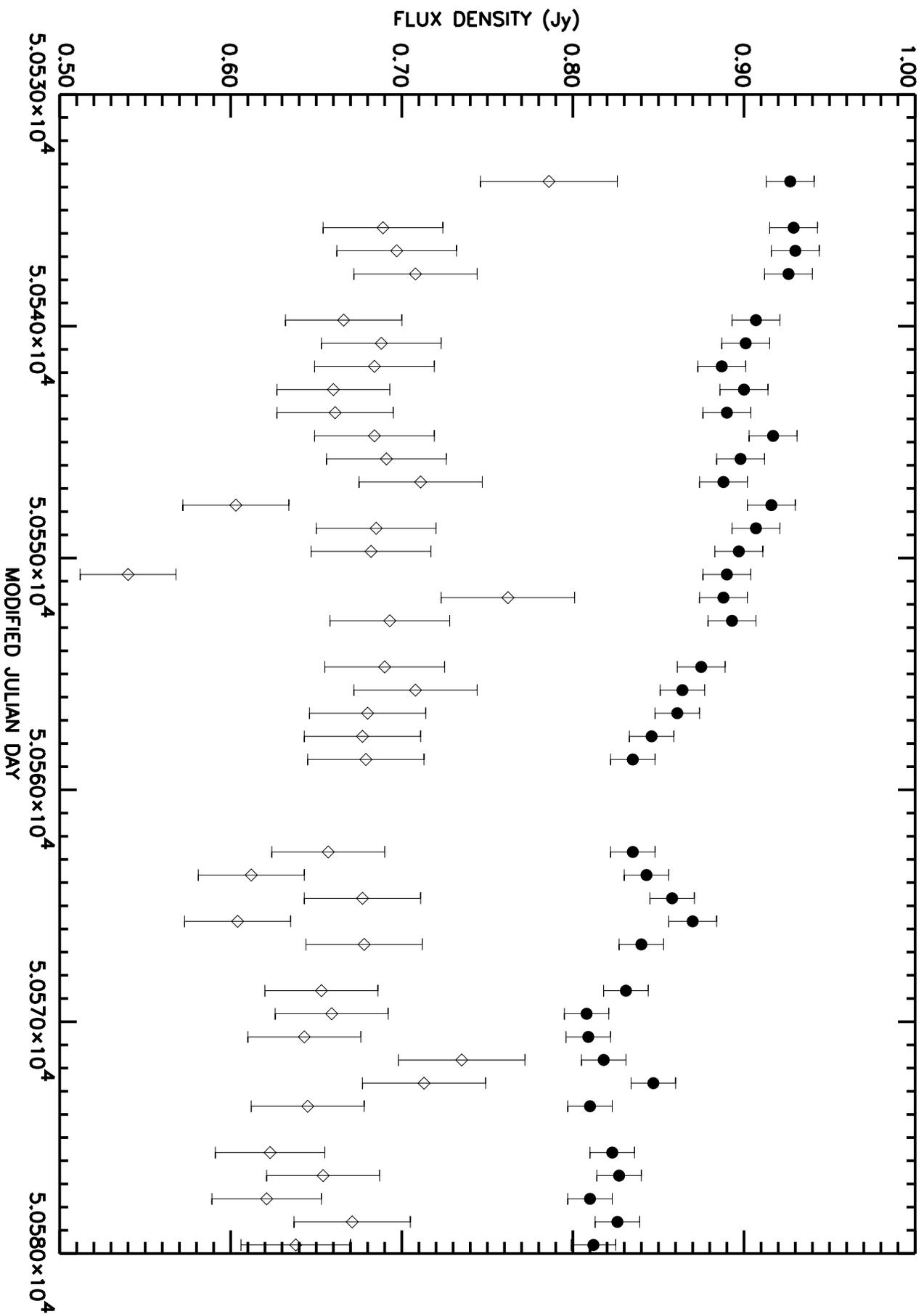
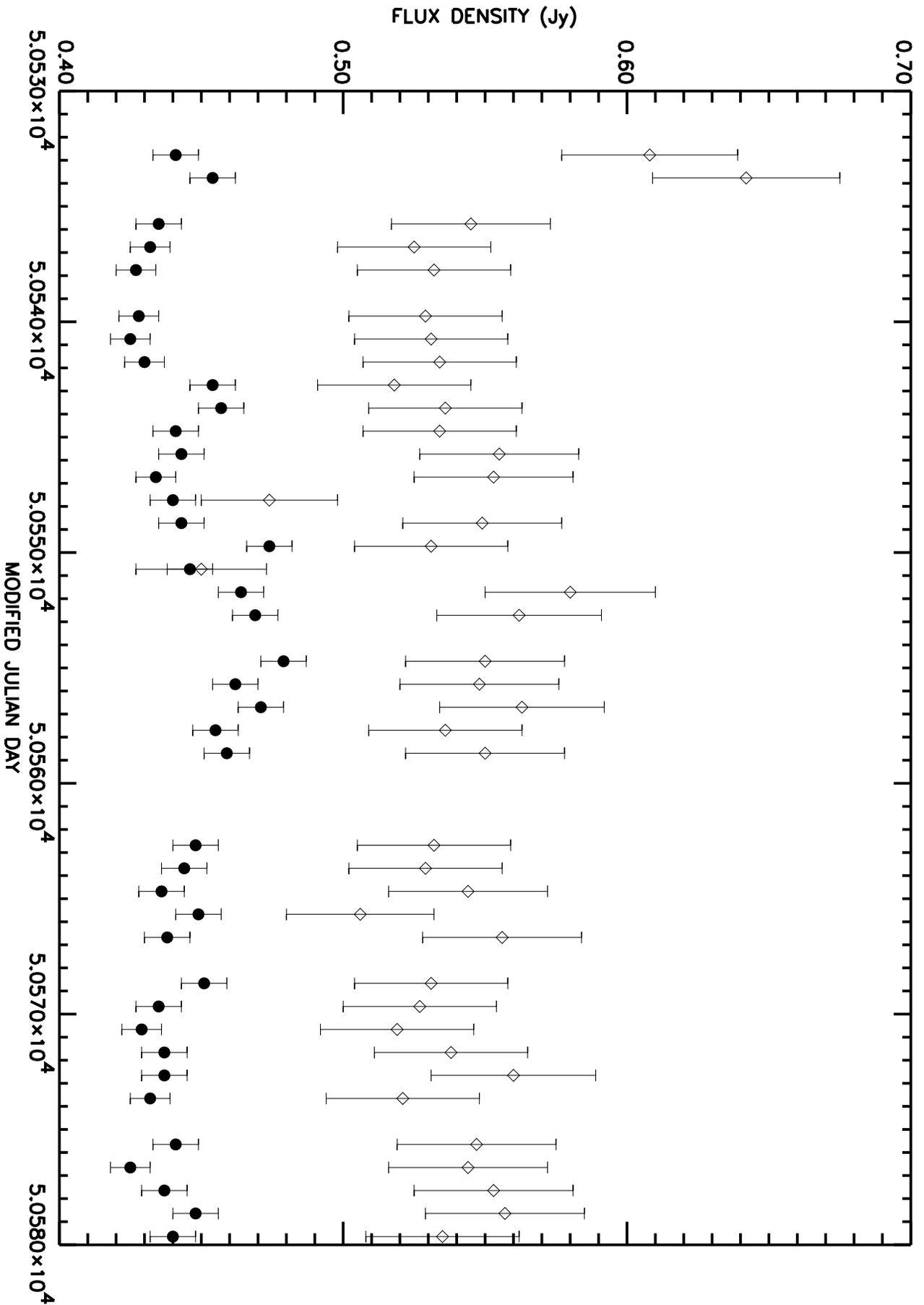
\end{document}